\documentclass[12pt]{article}

\usepackage[sort&compress,square,comma,authoryear]{natbib}
\usepackage{graphicx}
\usepackage{caption}
\usepackage{amsmath}
\usepackage{float}
\usepackage{subcaption}
\usepackage{physics}
\usepackage{authblk}
\usepackage{xcolor}

\begin{document}

\title{Oscillatory Residual Stresses in Steady Angular Channel Extrusion}
\author[1]{Arunava Ray}
\author[2,*]{Pritam Chakraborty}
\author[1]{Anindya Chatterjee}
\affil[1]{Mechanical Engineering, Indian Institute of Technology Kanpur, India}
\affil[2]{Aerospace Engineering, Indian Institute of Technology Kanpur, India}
\affil[*]{Corresponding author: Address: 210/F NWTF, Indian Institute of Technology Kanpur, UP-208016, India. email: cpritam@iitk.ac.in}
\date{}

\maketitle

\newpage

\begin{abstract}

Angular channel extrusion has evolved as processes that can induce significant strengthening of the formed product through grain refinement. However, significant residual stresses are developed in the extruded product whose quantification is necessary for accurate process design and subsequent heat treatment. Experimental evaluation of residual stress provides the through thickness (normal) variation at chosen sampling points on the formed product and may provide inaccurate estimates if variations along the extrusion (longitudinal) directions are present. Process models can complement the experimental measurements and improve the estimates of residual stress distribution. While models of this process have been developed, very few of them have been applied to understand the variation of residual stress in the formed products. The present work aims to address this limitation by providing a complete map of residual stress distribution in angular extrusion process through numerical simulations. Interestingly, our simulations show that the angular channel extruded product can have significant longitudinal variation of residual stress depending on the extrusion ratio and strain hardening rate. Detailed analyses of the process reveals that these spatial oscillations occur due to cyclic movement of the contact location between the die and the top-billet surface in the exit channel. The outcome of this study suggests that accurate measurement technique of residual stress field in angular channel extruded products should consider the possibility of longitudinal variations. The findings can be extended to other continuous forming processes as well.

\end{abstract}


\section{Introduction}
\label{sec:intro}

Angular channel extrusion is a manufacturing process where the workpiece is pressed through a die consisting of two channels intersecting at some angle \citep{Segal1981}. The process is associated with severe plastic deformation leading to grain refinement and subsequent strengthening of the formed product. However, as with other forming processes, significant residual stresses are developed for this process as well, which necessitate appropriate characterization to ensure that the performance of the formed products are not detrimentally affected by these locked-in stresses \citep{Schajer}. Contour method and X-ray diffraction technique have been employed in \cite{Khanlari12020}, and \cite{Romero2020}, respectively, to quantify the through thickness variation of residual stress in angular channel extruded products. However, owing to the complexity of these methods, variation of residual stress along the thickness of the formed product at few chosen locations could only be obtained. Typically full-field experimental measurement of residual stress is extremely difficult and the through thickness variations at certain sampling points are usually accepted as representative. Such a consideration is accurate if variations in the extrusion (longitudinal) directions are absent. However, these variations may be present depending on the process and material parameters. Thus, full-field map of residual stress is necessary to disallow any error due to oversight and numerical analyses can augment the experimental measurements in this respect.

Various analytical and numerical models of angular channel extrusion have been developed over the years. In the earliest work, \cite{Segal} gave an analytical estimation of the total strain as a function of the die angle of a test sample which was pressed through a die containing two equal channels. Frictionless contact between die and billet, fully filled die channel by workpiece, and sharp corners were assumed in the model. \cite{Iwahashi} extended the model proposed by \cite{Segal} for a multiple pass angular channel extrusion process. Finite Element Method (FEM) analyses (FEA) of the process by assuming plane strain conditions and considering the effect of friction was first presented by \cite{Prangell}. It was shown that with the inclusion of friction, the FEM predictions are in agreement with the upper bound solution. \cite{DeLo} presented 2-D and 3-D non-isothermal FEA of the process. The results, such as load versus stroke length of the punch, were compared with experiments for a Ti-6Al-4V alloy and a reasonable agreement was obtained. \cite{Kim1} used two different material models (strain hardening and perfectly plastic) with nearly equal yield stress values to explain the die-billet gap formation at the outer corner radius of the process using FEA. The gap between the billet and outer die radius was found larger for higher strain rate sensitive materials due to the relatively softer outer part. From the FEA, only the von Mises stress at the steady state deformed zone were presented. \cite{Kim}, \cite{Kim2} and \cite{Li} separately showed the effect of outer corner radius and friction on the extrusion load and effective plastic strain. \cite{Nagasekhar} simulated the angular channel extrusion process in ABAQUS/Explicit \citep{abaqus} for different friction conditions and showed that the extrusion load versus displacement curve from analysis have the same trend as with experiments on a copper billet. \cite{Leea} showed that the residual stress distribution from 3-D FEM simulations in ABAQUS was in good agreement with the experimental measurement using neutron diffraction. They  reported the radial variation of residual stresses, but the variation of the residual stress along longitudinal direction (LD) was not studied.

As is evident from the above review, a majority of the modeling work related to angular channel extrusion has been focused on understanding the effect of various process parameters on the deformation of the extruded product. Limited research activity can be found on residual stress analyses, and these only provide comparative studies on the variation along the through thickness or normal direction (ND). In the present work, 2D plane strain FEA of quasi-static angular extrusion for different die geometries and material properties is performed to investigate the variation of residual stresses in both the ND and LD. Interestingly, spatially oscillating residual stress profiles are clearly observed in the extruded component after single pass angular extrusion. Depending on the process parameters (such as extrusion ratio and hardening constants), the variation of the normal stress along the LD of the extruded billet can have either periodic or aperiodic oscillations, or no oscillations at all. The observed longitudinal variations can be physically related to the movement of the contact location between the die and billet in the exit section. These observations of oscillatory fields bring forth new possibilities of longitudinal variations of residual stresses in angular channel extrusion and other continuous forming processes, which so far has been assumed to be absent.

The organization of the paper is as follows: In section~\ref{sec:fea}, the finite element model of the angular extrusion process is presented. The details of the mesh sensitivity study using different values of ER are also shown. In section~\ref{sec:param_study}, the results of the parametric study using different values of ER and strain hardening rate are shown. The residual stress distribution for the different parameters is discussed in detail. The cause of residual stress variation along LD as observed in section~\ref{sec:param_study} is explained in detail in section~\ref{sec:discuss}. The conclusions from this study is presented in section~\ref{sec:conclu}.

\section{Numerical Model of the Angular Extrusion Process}
\label{sec:fea}

The details of the FEM model of the angular extrusion process and the elasto-plastic material parameters are presented in this section. Furthermore, the procedure to determine the size of elements that provides convergent solution is described.

\subsection{FEM Model and Material Parameters}
\label{sec:fea_props}
Two dimensional plane strain simulations of the angular extrusion process are carried out in the FEM software ABAQUS. An isotropic elasto-plastic material model with linear strain hardening has been used in the simulations.  The linear strain hardening response is a simplistic representation of stage III hardening in most metals/alloys and thus considered in the study. The hardening rate has been varied in the parametric study to analyze its influence on residual stress distribution. The values of Young's Modulus (E), Poisson's ratio ($\nu$) and yield stress are chosen as 200GPa, 0.3, and, 400 MPa, respectively, and kept constant in all the simulations.

\begin{figure}[H]
\centering
\includegraphics[width=0.4\textwidth]{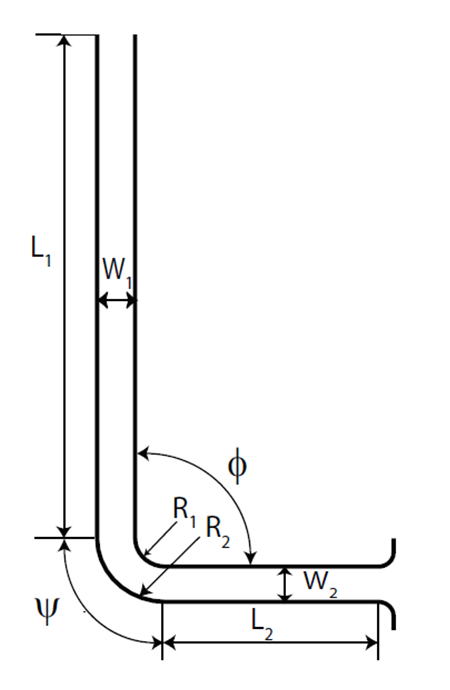}
\caption{Schematic of the die. The extrusion ratio ER=W$_2$/W$_1$.}
\label{fig:die_param}
\end{figure}

The schematic of the die geometry used in the simulations is shown in Fig.~\ref{fig:die_param}. In the figure, $W_1$ and $W_2$ are the widths of the inlet and the exit channel, respectively, with ER defined as $W_2$/$W_1$. The corner angle ($\Psi$) and die angle ($\phi$), as shown in the figure, are both $90^{\circ}$. $R_1$ and $R_2$ are the corner radii of the right and left wall of the die, respectively, while $L_1$ and $L_2$ are the lengths of entry and the exit channels, respectively. In this study the ER is varied by changing $W_2$ and keeping $W_1$ fixed at 0.025 m. All other geometric parameters are normalized with respect to $W_1$ and kept constant in the simulations. These parameters are shown in Table~\ref{table:die_param}.

\begin{table}[H]
\centering
\begin{tabular}{|c|c|c|c|c|c|}
\hline
Geometric parameters & $L_1$ & $L_2$ & $W_2$ & $R_1$ & $R_2$\\
\hline
Normalized value with respect to $W_1$ & 22 & 6 & ER & 0.8 & 1.8\\
\hline
\end{tabular}
\caption{Normalized geometric parameters of the die.}
\label{table:die_param}
\end{table}

In all the simulations, both the die and punch are considered rigid. The deformable billet is extruded through the die by first displacing the punch downwards with a constant velocity of 0.005 m/s through the vertical part. Later, when the bend is reached, points on the left face of the billet are pushed along appropriate arcs of circle until the bend has been fully traversed. Finally, the billet is pulled gently out using forces on the right face (this last phase is accompanied by very small deformations, because frictionless contact between the die, billet and punch is used). The billet is discretized using 4 node bilinear plane strain quadrilateral elements with reduced integration and hourglass control.

\subsection{Convergence Analysis of Mesh}
\label{sec:fea_mesh}

In the FEM simulations, the initial rectangular billet geometry is discretized using equisized 4-node quadrilateral elements. A sufficiently small element size is needed for accurately capturing the spatial gradients of the displacement and stress fields during the extrusion process. An $h$-refinement strategy is adopted whereby the domain is successively discretized with elements of the same family but of smaller size, followed by comparing responses along a chosen material line. In the absence of singularities, the differences between successive solutions should decrease with the reduction in element size. Once an acceptably small difference between the two solutions is obtained, the solutions are deemed to have converged and the coarser of the last two meshes is selected for subsequent simulations.

\begin{figure}[H]
\centering
\includegraphics[width=\textwidth]{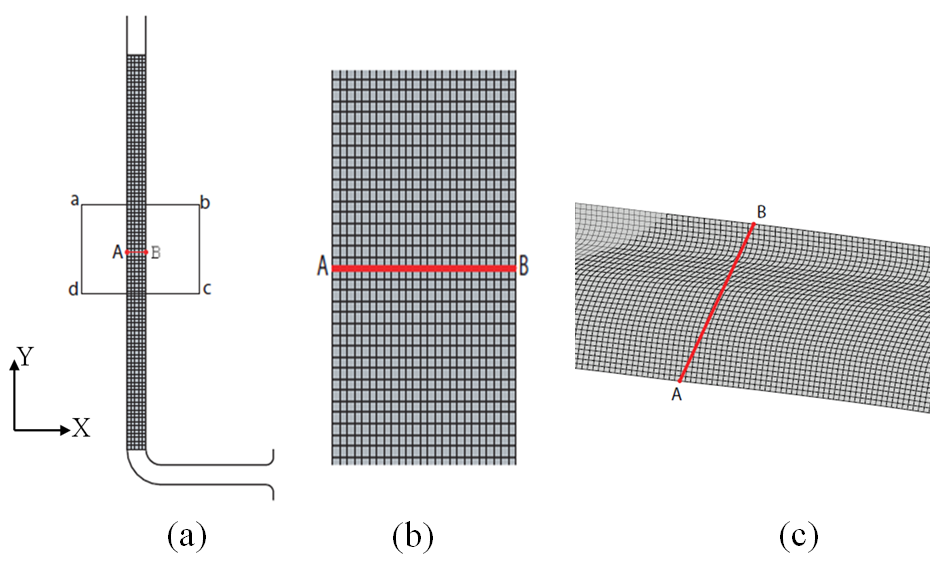}
\caption{(a) Initial configuration of the angular extrusion process. (b) Magnified view showing points A and B. (c) Magnified view of the final deformed configuration showing the material line on which $\sigma_{xx}$ is obtained.}
\label{fig:mesh_conv_path}
\end{figure}

Using the above strategy, the simulation cases 1 and 7 (the extreme cases within Table~\ref{table:fea_param_er}) are considered to determine the appropriate element size. FEM simulations are performed using three different element sizes of 1 mm, 0.758 mm and 0.5 mm, with the aspect ratio about 1 in each case. Local convergence of the computed solution is determined by comparing the interpolated nodal values of $\sigma_{xx}$ along a particular material line. To construct that material line, two points A and B are selected in the undeformed configuration of the billet as shown in Figs.~\ref{fig:mesh_conv_path}(a) and \ref{fig:mesh_conv_path}(b). The nodal $\sigma_{xx}$ values are extracted from material points on the straight line joining points A and B in the final deformed configuration (at the end of the simulation) as shown in Figure~\ref{fig:mesh_conv_path}(c). 

\begin{table}[H]
\centering
\begin{tabular}{|c|c|c|c|c|c|c|c|}
\hline
Case No.&1&2&3&4&5&6&7\\
\hline
Extrusion Ratio (ER) &1&0.9&0.8&0.7&0.75&0.6&0.5\\
\hline
\end{tabular}
\caption{The simulation cases considered in the FEA where the hardening rate is fixed at 5 MPa and the extrusion ratio is varied.}
\label{table:fea_param_er}
\end{table}

\begin{figure}[H]
\centering
\includegraphics[width=\textwidth]{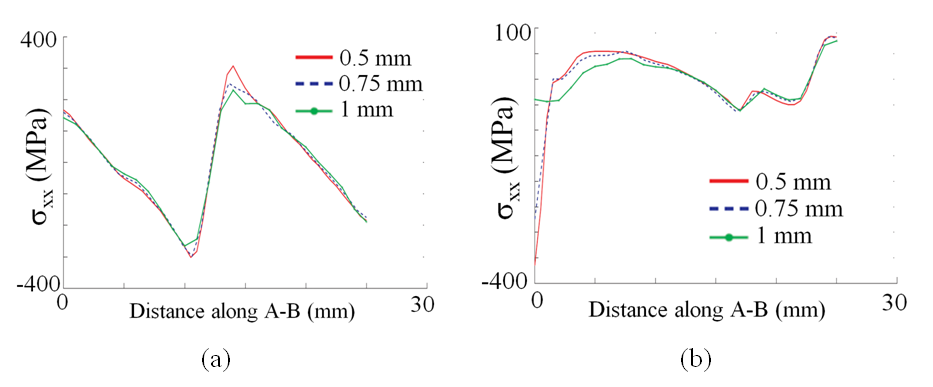}
\caption{Comparison of $\sigma_{xx}$ along material line shown in Fig.~\ref{fig:mesh_conv_path}(c) for three mesh sizes (0.5 mm, 0.758 mm, 1 mm) and simulation cases (a) 7 and (b) 1 (from Table~\ref{table:fea_param_er}).}
\label{fig:mesh_conv_s11}
\end{figure}

Results of the comparison are shown in Fig.~\ref{fig:mesh_conv_s11}. For simulation case 7 (Fig.~\ref{fig:mesh_conv_s11}(a)), the stress variations are almost identical along A-B. However, for case 1 (Fig.~\ref{fig:mesh_conv_s11}(b)), a deviation can be observed near point A for the coarsest mesh, while results are almost overlapping for element sizes 0.758 and 0.5 mm, respectively. Based on this analysis, it can be concluded that the element size of 0.758 mm provides a reliable solution and hence this size is used in all subsequent simulations.

\section{Parametric Study and Distribution of Residual Stress}
\label{sec:param_study}

A parametric study is now reported wherein the extrusion ratio (ER) and the hardening rate are varied. In the first seven simulation cases the hardening rate is kept constant at 5 MPa and the ER is varied from 1 to 0.5 (Table~\ref{table:fea_param_er}). The variation of residual stress for these cases are shown in section~\ref{sec:er_sensitive}. In the next seven simulation cases, the ER is kept constant at 0.75 and the hardening rate is varied (Table~\ref{table:fea_param_hard_rate}). The results of these cases are discussed in section~\ref{sec:st_rate_sensitive}. Specifically the stress component $\sigma_{xx}$ along both the ND and LD are presented for the different simulation cases.

\begin{table}[H]
\centering
\begin{tabular}{|c|c|c|c|c|c|c|c|}
\hline
Case No.&8&9&10&11&12&13&14\\
\hline
Hardening Rate (MPa)&20&60&160&260&360&460&560\\
\hline
\end{tabular}
\caption{Simulation cases considered in the FEA where the ER is fixed at 0.75 and the hardening rate is varied.}
\label{table:fea_param_hard_rate}
\end{table}

\begin{figure}[H]
\centering
\includegraphics[width=0.4\textwidth]{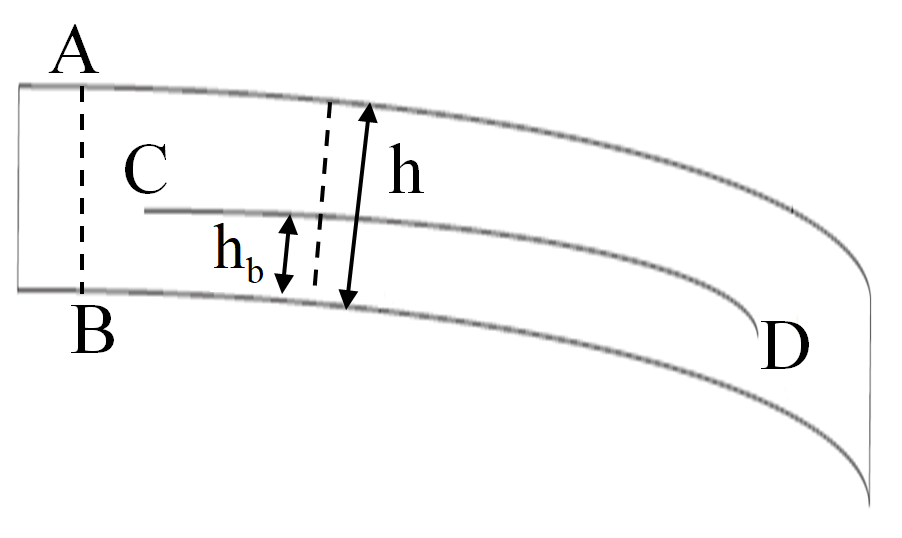}
\caption{A region of the extruded billet showing a representative section A-B on which $\sigma_{xx}$ is evaluated, and curve C-D obtained by joining all the points with maximum sectional $\sigma_{xx}$.}
\label{fig:linexx}
\end{figure}

The variation of $\sigma_{xx}$ along ND is obtained at multiple sections along the length of the extruded billet. These sections are constructed such that they are approximately perpendicular to the top and bottom surface of the billet (Fig.~\ref{fig:linexx}). The values of $\sigma_{xx}$ are then interpolated on these sections from the Gauss-point values. Fig.~\ref{fig:linexx} shows a typical extruded 2D billet along with the sectional line A-B used to extract the ND variation of $\sigma_{xx}$.

Variations of $\sigma_{xx}$ along LD are then evaluated from several sections like A-B separated by $\approx$ 1 mm, along the length of the extruded billet. The maximum value of $\sigma_{xx}$ is first evaluated on all these sections and the variations along LD are then reported on curve C-D that connects all these points as shown in Fig.~\ref{fig:linexx}. For all the simulation cases, the curve CD remains nearly parallel to the bottom surface of the extruded billet. However, the distance of the curve from the bottom surface ($h_b$) changes with ER and is shown in Table~\ref{table:linexx}. From the table it can be observed that the maximum $\sigma_{xx}$ is below the medial axis ($h_b/h$=0.5) for ER $\ge$ 0.75 and is followed by a sudden shift around ER = 0.7. This also signifies that there is a flipping of the residual stress profile along ND as ER is lowered below about 0.75. The profiles below ER = 0.7 are also qualitatively similar to that obtained in elasto-plastic bending \citep{Joudaki2015}.

\begin{table}[H]
\centering
\begin{tabular}{|c|c|c|}
\hline
ER & $h$ (m) & $h_b/h$ \\
\hline
1.0&0.02369&0.40805\\
0.9&0.02157&0.3615\\
0.8&0.01935&0.3223\\
0.75&0.01825&0.2445\\ 
0.7&0.01711&0.5780\\
0.6&0.01480&0.5149\\
0.5&0.01239&0.5451\\
\hline
\end{tabular}
\caption{Ratio of $h_b$ to $h$ (see Fig.~\ref{fig:linexx}) for different ERs.}
\label{table:linexx}
\end{table}

\subsection{Effect of Extrusion Ratio (ER)}
\label{sec:er_sensitive}

\begin{figure}[H]
\centering
\includegraphics[width=0.7\textwidth]{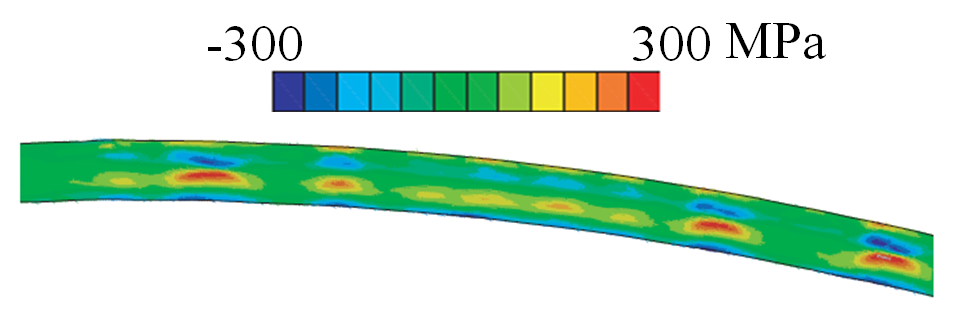}
\caption{Residual stress component $\sigma_{xx}$ in the extruded billet for ER = 1.0.}
\label{fig:s11_cont_aperiodic}
\end{figure}

\begin{figure}[H]
\centering
\includegraphics[width=\textwidth]{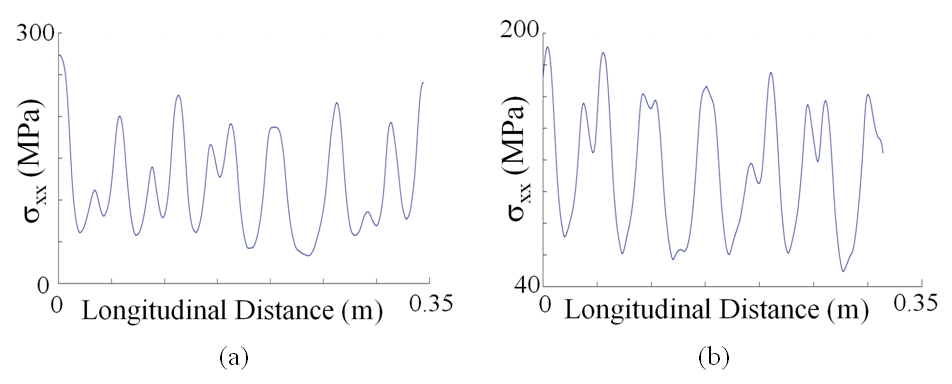}
\caption{Maximum of residual stress component $\sigma_{xx}$ on every section perpendicular to the medial axis along the LD of the extruded billet for (a)
ER = 0.9 and (b) ER = 1.0.}
\label{fig:s11_curve_aperiodic}
\end{figure}

\begin{figure}[H]
\centering
\includegraphics[width=0.7\textwidth]{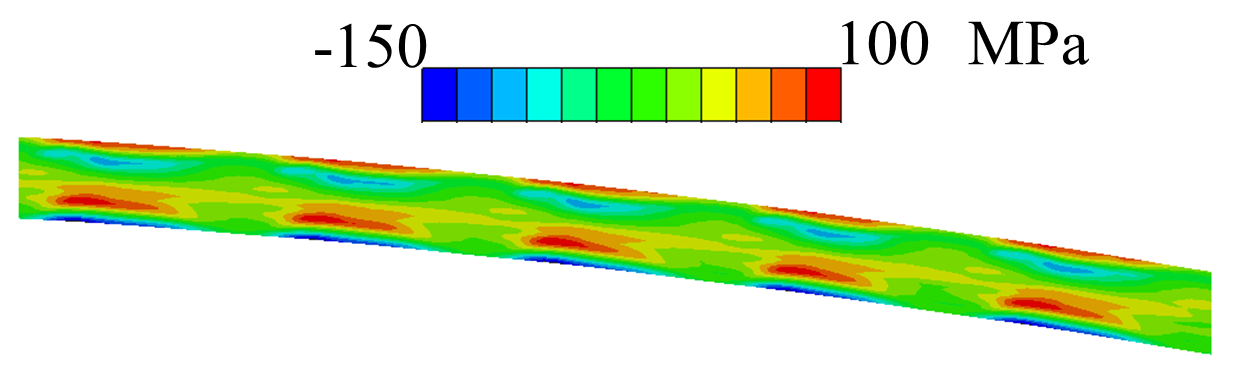}
\caption{Residual stress component $\sigma_{xx}$ in the extruded billet for ER = 0.75.}
\label{fig:s11_cont_periodic}
\end{figure}

\begin{figure}[H]
\centering
\includegraphics[width=\textwidth]{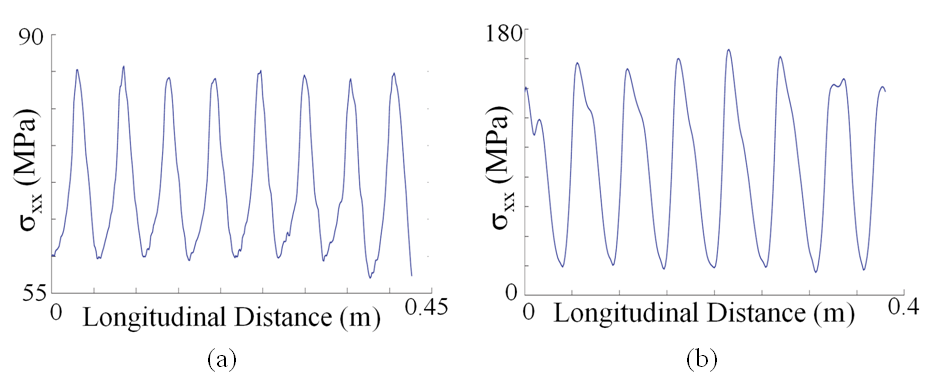}
\caption{Maximum residual stress component $\sigma_{xx}$ on successive sections perpendicular to the medial axis along the LD of the extruded billet for (a) ER = 0.7 and (b) ER = 0.8.}
\label{fig:s11_curve_periodic}
\end{figure}

\begin{figure}[H]
\centering
\includegraphics[width=0.8\textwidth]{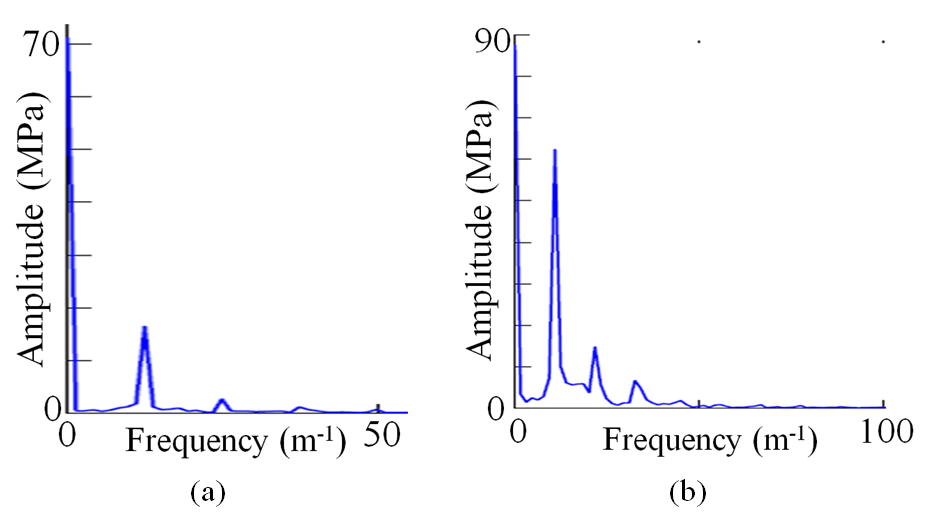}
\caption{Fast Fourier Transform (FFT) of the responses shown in (a) Fig.~\ref{fig:s11_curve_periodic}(a) and (b) Fig.~\ref{fig:s11_curve_periodic}(b).}
\label{fig:s11_fft_periodic}
\end{figure}

As seen in Figs.~\ref{fig:s11_cont_aperiodic} and \ref{fig:s11_curve_aperiodic}, for ER =  1 and 0.9, there is clearly aperiodic longitudinal variation of maximum $\sigma_{xx}$. For ER =  0.8 and lower (until 0.7), the variation is much closer to periodic, as seen in Figs.~\ref{fig:s11_cont_periodic} and \ref{fig:s11_curve_periodic}. A Fourier transform of the longitudinally varying maximum sectional $\sigma_{xx}$ (Fig.~\ref{fig:s11_fft_periodic}) shows the presence of a dominant frequency consistent with the nearly periodic behavior.

\begin{figure}[H]
\centering
\includegraphics[width=0.7\textwidth]{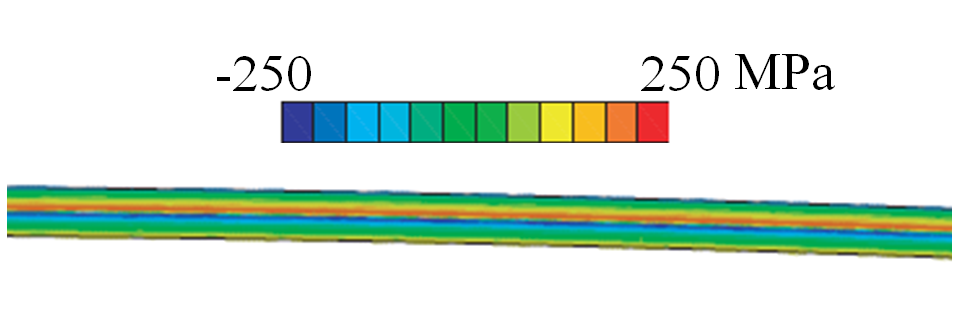}
\caption{Residual stress component $\sigma_{xx}$ in the extruded billet for ER = 0.6.}
\label{fig:s11_cont_steady}
\end{figure}

\begin{figure}[H]
\centering
\includegraphics[width=0.5\textwidth]{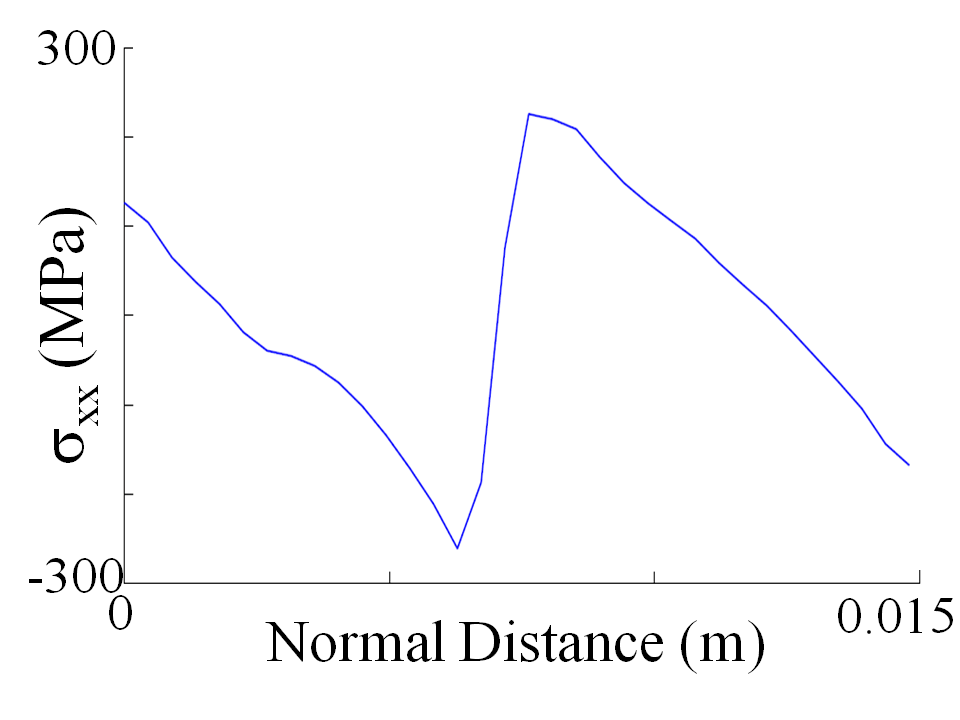}
\caption{Residual stress component $\sigma_{xx}$ in the ND for ER = 0.6.}
\label{fig:s11_normal_curve_steady}
\end{figure}

As the ER is reduced further to 0.6, the oscillation in the residual stress field disappears, as seen in Fig.~\ref{fig:s11_cont_steady}. The through thickness variation of $\sigma_{xx}$, as shown in Fig.~\ref{fig:s11_normal_curve_steady}, displays the familiar zigzag pattern developed due to plastic bending and subsequent elastic unloading \citep{Joudaki2015}.

\begin{figure}[H]
\centering
\includegraphics[width=0.5\textwidth]{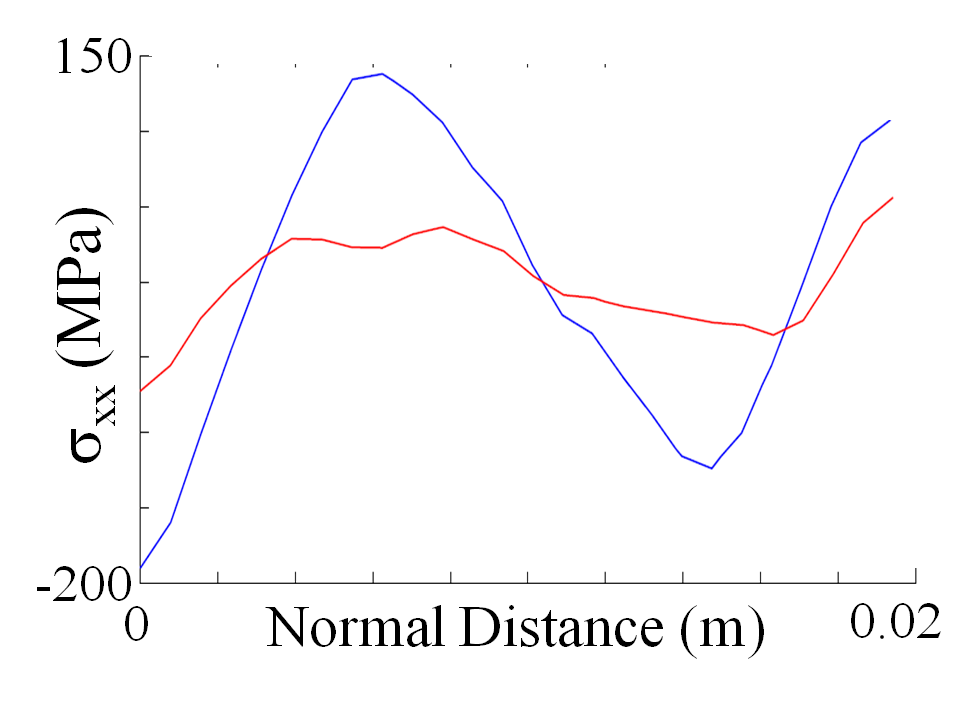}
\caption{Residual stress component $\sigma_{xx}$ in the ND for ER = 0.8 at 2 different perpendicular sections along the LD.}
\label{fig:s11_normal_curve_0_80}
\end{figure}

\begin{figure}[H]
\centering
\includegraphics[width=\textwidth]{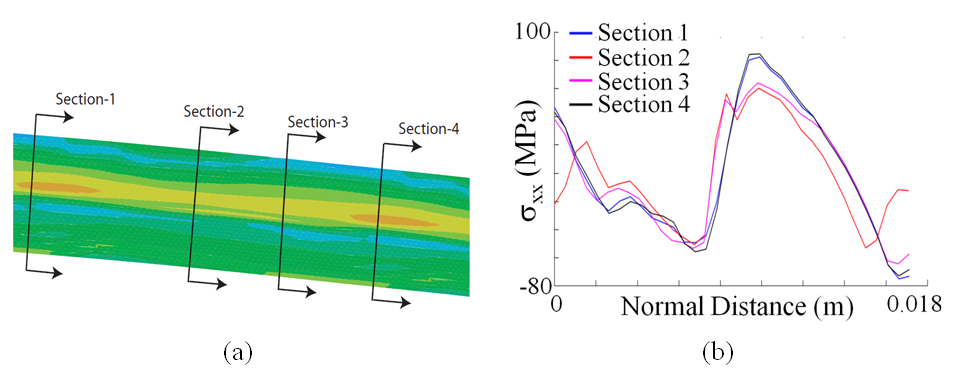}
\caption{Through thickness variation of $\sigma_{xx}$ for ER = 0.7. (a) Sections perpendicular to medial axis on which the variation of $\sigma_{xx}$ along the ND is compared. (b) Variation of $\sigma_{xx}$ along the ND.}
\label{fig:s11_normal_var_0_70}
\end{figure}

Interestingly, the residual stress distribution along the ND for higher ERs displays a sense opposite to that in Fig.\ \ref{fig:s11_normal_curve_steady}, as seen in Fig.~\ref{fig:s11_normal_curve_0_80}. The transition of the stress distribution profile begins to occur near ER = 0.7 whereby for the different sections along the extrusion direction (Fig.~\ref{fig:s11_normal_var_0_70}(a)), the residual stress distribution along the ND starts to change sign (Fig.~\ref{fig:s11_normal_var_0_70}(b)).

\subsection{Effect of Strain Hardening Rate}
\label{sec:st_rate_sensitive}

\begin{figure}[H]
\centering
\includegraphics[width=0.7\textwidth]{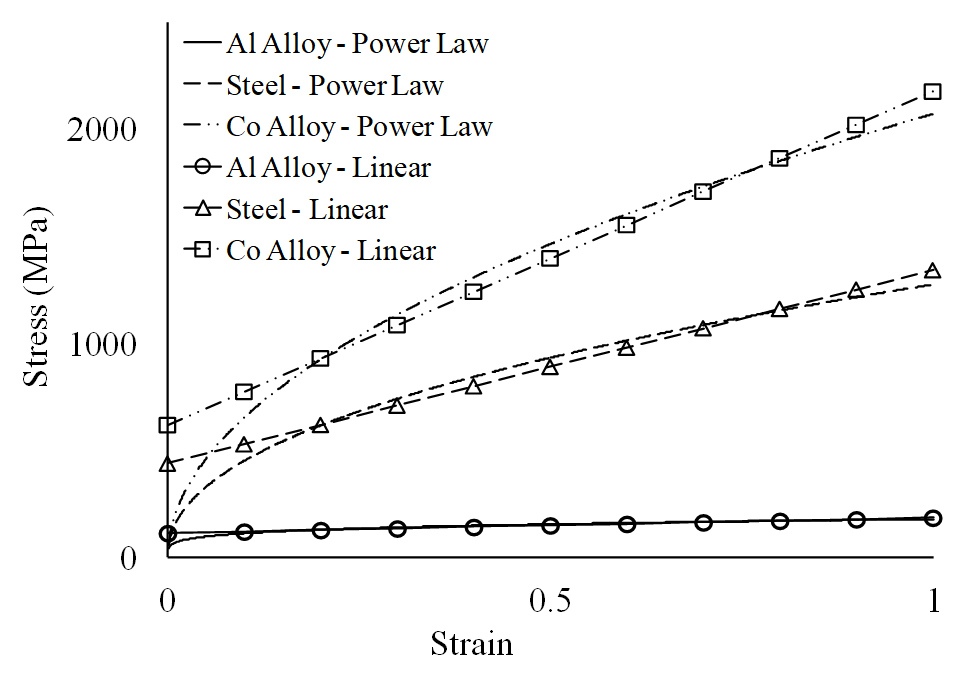}
\caption{Linear fits of power law hardening curves of Aluminum and Cobalt alloy, and stainless steel at room temperature. The fits are calculated using the stress values obtained from the power law for strain values from 5 to 100 \%.}
\label{fig:hardening_fit}
\end{figure}

In the previous section, results are presented for different ERs with a fixed hardening rate of 5 MPa as shown in Table~\ref{table:fea_param_er}. In this section, the ER is fixed at 0.75 and the hardening rate is increased from 5 MPa to 560 MPa (Table~\ref{table:fea_param_hard_rate}) to analyze its influence on residual stress. The strain hardening behavior for most metals and alloys can be suitably represented by a power law \citep{Hosford2007}. However, in this work a linear fit is considered since it provides a reasonable approximation of the stress-strain response beyond the initial portion of the curve. The linear fits of the power law curves for Aluminum alloy, Cobalt alloy, and annealed stainless steel at room temperature are shown in Fig.~\ref{fig:hardening_fit}. The parameters of the power law model for these three alloys are taken from \cite{Callister2005} and \cite{Kalpakjian2014}. The linear hardening rates are obtained as 70, 900 and 1500 MPa for Aluminum alloy, stainless steel and Cobalt alloy, respectively. Thus, linear hardening rates of approximately 70 MPa and higher are representative of room temperature response of metals and alloys while lower rates are more representative of hardening behavior at elevated temperatures.

\begin{figure}[H]
\centering
\includegraphics[width=\textwidth]{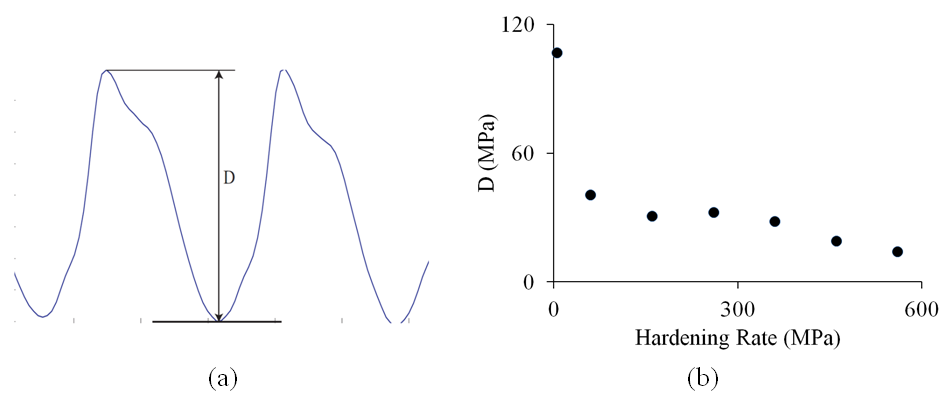}
\caption{(a) The variable 'D', peak to peak amplitude of nearly periodic oscillation of maximum sectional $\sigma_{xx}$ along the LD. (b) Variation of 'D' with strain hardening rate.}
\label{fig:d_param}
\end{figure}

Consistent with observations made in section \ref{sec:er_sensitive}, nearly periodic oscillation along the LD is observed for ER of 0.75 and hardening rate of 5 MPa, and a magnified view of the same is shown in Fig.~\ref{fig:d_param}(a). A similar longitudinal variation of residual stress is obtained from simulations for the other hardening rates at ER = 0.75. However, a decrease in the amplitude of oscillations is observed with increase in hardening rate. To quantify this observation, a variable `$D$' is defined, which is the difference between the maximum and minimum of the peak sectional $\sigma_{xx}$, along the longitudinal direction (LD) and over one cycle, as shown in Fig.~\ref{fig:d_param}(a). For a particular hardening rate at ER = 0.75, $D$ remains roughly constant from cycle to cycle. Hence, $D$ can be considered as twice the amplitude of the oscillation. A comparison of $D$ for different hardening rates is shown in Fig.~\ref{fig:d_param}(b), and clearly reveals a decrease in oscillation amplitudes with the increase in hardening rate. Thus, hot extrusion can potentially lead to larger oscillations in commonly used metals and alloys than cold extrusion. Correspondingly, cold extrusion of steel is expected to generate comparatively smaller oscillations in residual stress when compared to aluminium, since the former has a higher hardening rate than the latter.

\section{Role of Contact Conditions in Residual Stress Variation}
\label{sec:discuss}

Our simulations show that details of the contact interaction between the billet and the exit channel play a strong role in the creation of oscillatory residual stresses. During extrusion, the width of the billet actually becomes slightly less than that of the die exit section. Immediately past the bend, contact is steadily maintained between the billet and the lower surface of the exit section; however, due to the billet being slightly narrower, contact is lost at the upper surface. Subsequently, as the billet moves along the exit section of the die, there is a small upward curvature in the billet which causes loss of contact at the lower die surface. The degree of upward curvature is found to vary with time, leading to a sequence of fresh contact points being established between the billet and the upper die surface. Each successive contact point, thus formed, subsequently moves downstream along with the billet. When a contact point has moved sufficiently far downstream, the curvature in the billet increases again, and a new contact point is formed between the billet and the upper die surface. This mechanism of space- and time-varying contact force distribution causes the appearance of oscillatory residual stresses. The rest of this section is devoted to a detailed discussion of the same.

We begin by comparing the temporal variation of the state of contact and the spatial variation of the residual stress field along the LD for different cases (Table~\ref{table:fea_param_er}). To this end, we fix a spatial region and examine
nodal values of $\sigma_{yy}$ on the top and bottom surfaces of the portion
of the billet that lies inside this region. Nodes are taken to be not in contact if their $\sigma_{yy}$= 0. Conversely, nonzero values of $\sigma_{yy}$ indicate
the presence of normal contact tractions. The results obtained are depicted graphically for several discrete instants of time, which shows the evolution
of the contact state.

\subsection{Nearly Periodic Oscillations in Residual Stress}
\label{sec:case5_contact}

\begin{figure}[H]
\centering
\includegraphics[width=0.6\textwidth]{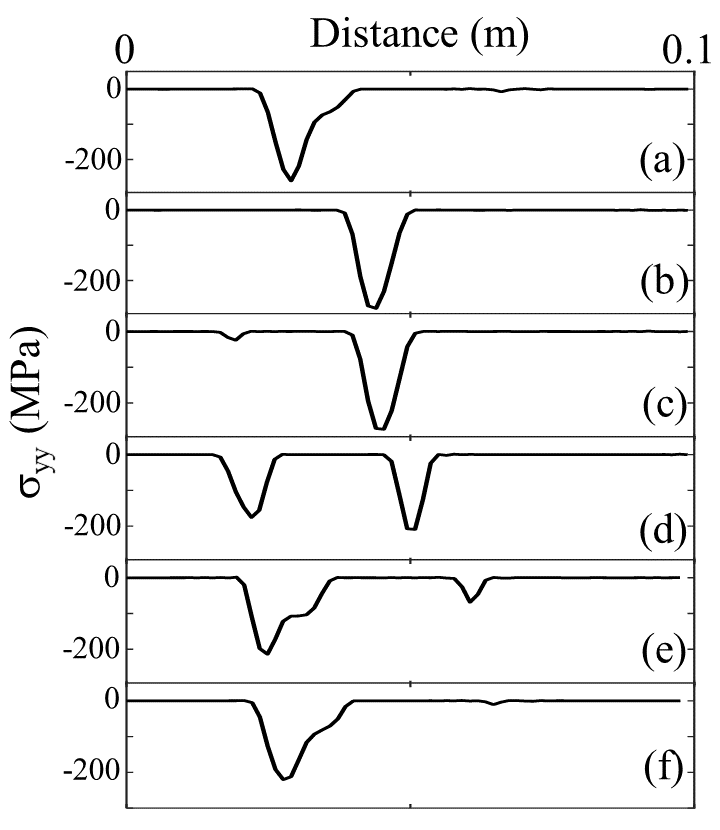}
\caption{Stress component $\sigma_{yy}$ on the top billet surface in the exit channel for case 5 (Table~\ref{table:fea_param_er}), at time instants (a) t$_0$ (an arbitrary starting time), (b) t$_0$ + 4.41 s, (c) t$_0$ + 4.48 s, (d) t$_0$ + 5.35 s, (e) t$_0$ + 7.05 s, (f) t$_0$ + 7.68 s.}
\label{fig:s22_er_0_75_time}
\end{figure}

From section \ref{sec:er_sensitive} it is observed that the variation of residual stress along the LD shows nearly periodic oscillation for case 5 (Table~\ref{table:fea_param_er}). For this case, nodal $\sigma_{yy}$ on the top surface of the billet inside the region of observation are shown at some instants of time in Fig.~\ref{fig:s22_er_0_75_time}. As seen from the plots, there is significant variation in the
contact state over time. In particular, a region of high contact traction moves to the right with the billet, until a new contact is established; subsequently, both contacts move downstream; as each contact moves downstream, its strength
initially increases and then decreases to zero. Comparing figures \ref{fig:s22_er_0_75_time}(a) and \ref{fig:s22_er_0_75_time}(f), we see that a nearly periodic variation is obtained. This nearly periodic behavior persisted until the billet was completely extruded past the bend. For the case shown in the figure, the time period obtained is 7.68 sec (approximately). The residual stress field in the extruded billet for this case is shown in Fig. \ref{fig:s11_cont_periodic} which shows nearly periodic oscillation along the LD.  

\begin{figure}[H]
\centering
\includegraphics[width=0.5\textwidth]{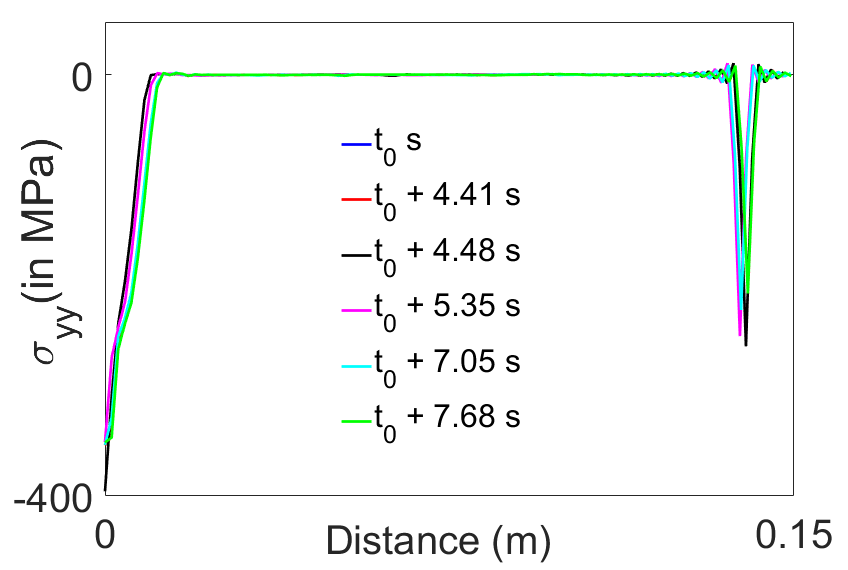}
\caption{Stress component $\sigma_{yy}$ at the bottom billet surface in the exit channel for case 5 (Table~\ref{table:fea_param_er}) at different instants of time.}
\label{fig:s22_bottom_er_0_75}
\end{figure}

In notable contrast to the traction distributions on the top surface, the contact state on the bottom surface of the billet remains essentially unchanged over the same time duration, as shown in Fig.~\ref{fig:s22_bottom_er_0_75}.

\subsection{Aperiodic Oscillations in Residual Stress}
\label{sec:case1_contact}

\begin{figure}[H]
\centering
\includegraphics[width=0.5\textwidth]{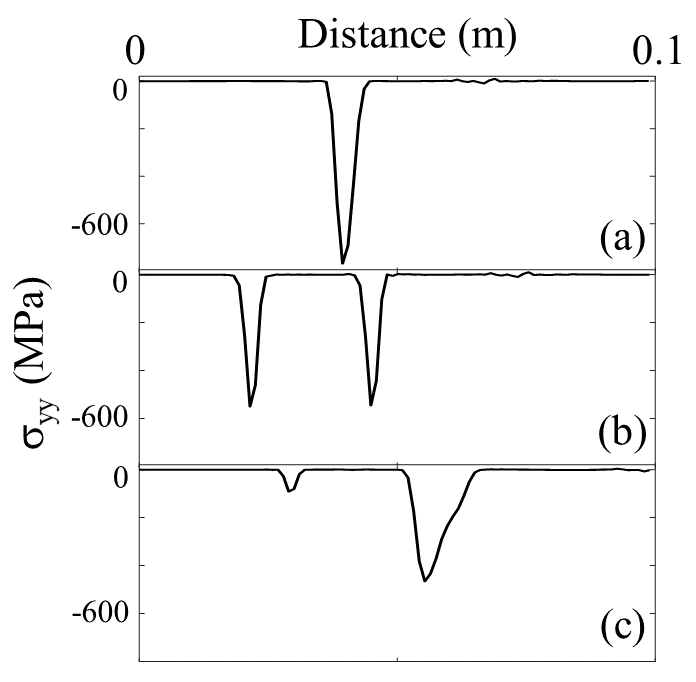}
\caption{Stress component $\sigma_{yy}$ at time instants (a) t$_0$ (an arbitrary starting time) (b) t$_0$ + 1.89 s (c) t$_0$ + 11.02 s, on the top billet surface in the exit channel for case 1.}
\label{fig:s22_er_1_00_time}
\end{figure}

We have observed from simulations that aperiodic temporal variation of contact results in aperiodic oscillation of residual stress in the billet along the LD. Some details are shown below for case 1 of Table~\ref{table:fea_param_er}. The temporal variation of contact tractions is shown in Fig.~\ref{fig:s22_er_1_00_time}, where no periodic pattern is seen. Thus, periodicity or aperiodicity in contact conditions between the billet and the upper section of the exit channel govern periodicity or aperiodicity of residual stress in the final billet. The residual stress field in the extruded billet for this case is shown in Fig. \ref{fig:s11_cont_aperiodic} which shows  aperiodic oscillation along the LD.

\subsection{Non-Oscillatory Residual Stress}
\label{sec:case6_contact}

\begin{figure}[H]
\centering
\includegraphics[width=0.6\textwidth]{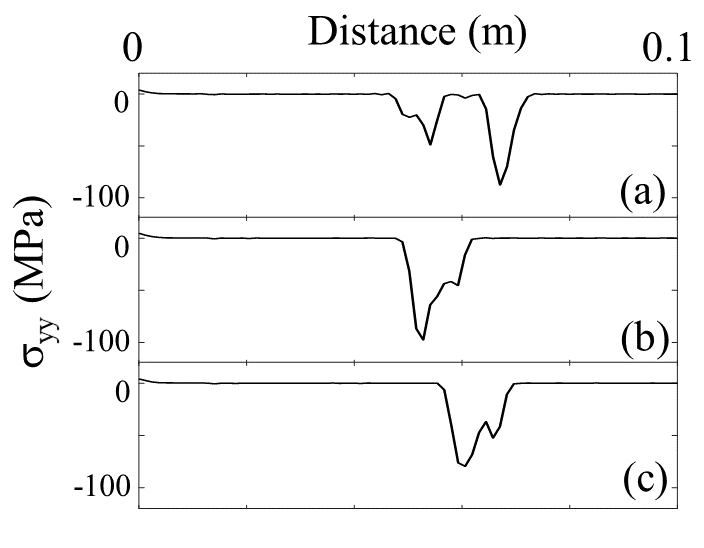}
\caption{Stress component $\sigma_{yy}$ at time instants (a) t$_0$ (an arbitrary starting time) (b) t$_0$ + 2.66 s (c) t$_0$ + 5.66 s, on the top billet surface in the exit channel for ER = 0.6.}
\label{fig:s22_er_0_60_time}
\end{figure}

The above mechanism also applies to situations where near zero oscillations in residual stress along the LD are observed. The temporal variation of contact condition for case 6 (Table~\ref{table:fea_param_er}) is shown in Fig.~\ref{fig:s22_er_0_60_time} as an example. In the figure, the location of the contact region remains nearly the same over time. Minor differences in traction distribution are observed, but they remain localized and have a small effect overall. The result is a longitudinally steady residual stress field as evident from Fig. \ref{fig:s11_cont_steady}.

\section{Conclusion and Discussion}
\label{sec:conclu}

Angular channel extrusion is a continuous forming process that leads to grain refinement and significant strengthening of the product. However, residual stresses can be developed in the formed products that may degrade their performance. Hence, a detailed quantification of the residual stress field of the extruded product needs to be obtained. Experimental measurement of residual stress is typically performed at few locations to obtain the variation along ND. Inferences of the developed residual stress field are then made based on these measurements and with the assumption that variations along LD are absent. However, the assumption need not be valid in general, and can depend on the material and process conditions.

In the present work, a numerical study of the angular channel extrusion process has been performed to explore the possibility of development of longitudinally varying residual stress. Plane strain large deformation elasto-plastic FEM simulations have been performed for different strain hardening rates and ER. The analyses show that at low strain hardening rate and high ER, significant oscillations of residual stress along LD can occur. A closer investigation reveals that a small separation between the billet with the die at the bend into the exit channel occurs due to the intense deformation in the bend. Subsequently, contact is restored as the billet travels along the exit channel. The contact location itself can move in space, either periodically or aperiodically. The distance of the contact location from the bend can be thought of as providing a moment arm for the contact force. This moment due to the contact forces influences the deformation process going on within the bend. The variation of contact location changes the applied moment which modifies the deformation behavior in the bend. These variable deformation histories cause sectionwise variable residual stresses, which are swept down the exit channel without further change and eventually emerge as spatially oscillating residual stresses in the billet.

The extent of variation of contact location in the exit channel is strongly dependent on the ER. A lower ER reduces the lengths of the regions in which loss and reestablishment of contact of the billet with the die at the exit channel can occur. Thus, variation of the moment on the deforming billet in the bend due to contact forces at the exit channel is reduced. As this moment influences the deformation behavior in the bend, a reduction in its variation lessens the changes in the history of deformation resulting in decreased variation of residual stress along LD. An increase in strain hardening rate reduces the sensitivity in the deformation behavior of the billet in the bend on the moment due to contact forces at the exit channel. This leads to a reduction in the variation of residual stress in the billet along LD for increasing strain hardening rate. Thus, for a given ER, if we increase the hardening {\em rate} then we expect the oscillatory residual stresses to disappear; and for a given hardening rate, if we reduce the ER, we expect the residual stresses to disappear as well. Both of these effects have been convincingly observed in the numerical simulations in this paper.

From this study it can be concluded that longitudinal oscillation of residual stress is possible in angular channel extrusion depending on the geometry parameters of the die and material parameters of the billet. This observation clearly indicates that enough sampling points should be considered while measuring residual stresses in formed products for the purpose of material characterization, process design and heat treatment. The observations can be extended to other continuous forming processes where variations on contact is possible.

\bibliography{references}
\bibliographystyle{apa}

\end{document}